\documentclass[12pt]{article}
\usepackage{amssymb}
\usepackage{amsmath}
\usepackage{amstext}
\usepackage{graphicx,epsfig}
\usepackage{epsfig}
\usepackage{verbatim} 
\usepackage{fancyhdr}
\usepackage{fancybox}
\usepackage{color}
\usepackage{ulem,bbold}
\usepackage{enumitem}
\usepackage{subfigure}
\usepackage{bbm}
\usepackage{parskip}
\usepackage{cite}

\newcommand{\Comment}[1]{{}}
\definecolor{MyDarkBlue}{rgb}{0.15,0.15,0.45}
\usepackage[linktocpage=true]{hyperref}
\hypersetup{
colorlinks=true,
citecolor=MyDarkBlue,
linkcolor=MyDarkBlue,
urlcolor=MyDarkBlue,
}

\newcommand{\be}{\begin{equation}}
\newcommand{\ee}{\end{equation}}
\newcommand{\bea}{\begin{eqnarray}}
\newcommand{\eea}{\end{eqnarray}}
\newcommand{\beas}{\begin{eqnarray*}}
\newcommand{\eeas}{\end{eqnarray*}}
\newcommand{\nn}{\nonumber}

\newcommand{\half}{\frac{1}{2}}

\numberwithin{equation}{section}

\usepackage{genyoungtabtikz}
\Yboxdim{13pt} 
\Ylinethick{.8pt}

\begin{document}


\begin{center}
{\Large \bf{Shift Symmetries for $p$-Forms and\\ }}
\vspace{.2cm}
{\Large \bf{Mixed Symmetry Fields on (A)dS}}
\end{center} 
 \vspace{1truecm}
\thispagestyle{empty} \centerline{
{\large {Kurt Hinterbichler}}$^{}$\footnote{E-mail: \Comment{\href{mailto:kurt.hinterbichler@case.edu}}{\tt kurt.hinterbichler@case.edu}} 
}

\vspace{1cm}

\centerline{{\it ${}^{}$CERCA, Department of Physics,}}
\centerline{{\it Case Western Reserve University, 10900 Euclid Ave, Cleveland, OH 44106}} 

 \vspace{1cm}

\begin{abstract} 

Massive fields on (anti) de Sitter space realize extended shift symmetries at particular values of their masses.  We find these symmetries for all bosonic $p$-forms and mixed symmetry fields, in arbitrary spacetime dimension.  These shift symmetric fields correspond to the missing longitudinal modes of mixed symmetry partially massless fields where the top row of the Young tableau is activated.  

\end{abstract}

\newpage

\thispagestyle{empty}
\tableofcontents

\setcounter{page}{1}
\setcounter{footnote}{0}

\parskip=5pt
\normalsize

\section{Introduction}

In \cite{Bonifacio:2018zex}, shift symmetries acting on massive symmetric tensor fields on (anti)-de Sitter space ((A)dS) were found, generalizing the extended shift symmetries of the galileon \cite{Luty:2003vm,Nicolis:2008in} and special galileon \cite{Cheung:2014dqa,Hinterbichler:2015pqa,Cheung:2016drk,Novotny:2016jkh} on flat space.  We refer to the introduction of \cite{Bonifacio:2018zex} for more extensive background and motivation.

A summary of the results of \cite{Bonifacio:2018zex} is as follows.  A massive spin $s$ field on AdS$_D$ of radius $L$, carried by a symmetric tensor field $\phi_{\mu_1\cdots\mu_s}$, satisfies on shell the Klein-Gordon equation,
\be \begin{cases} \left(\nabla^2 -m^2\right)\phi=0\,\, ,\ \ \ \ \ \ \ \ \ \ \ \ \ \ \ \ \ \ \ \ \ \ \ \ \ \ \ \ \ \ \ \ \ \ \ \ \ \ \ \ \ \ \ \ \ \ \ \ \ \ \ \ \ \ \ \ \ \ \ s=0\,, \\
\left(\nabla^2+{1\over L^2}\left[s+D-2-(s-1)(s+D-4)\right]-m^2\right)\phi_{\mu_1\cdots\mu_s} = 0\ ,\  \ s\geq 1\,,
\end{cases} \ee
along with transversality in all indices and full tracelessness.  (Here the mass squared $m^2$ for $s\geq1$ is chosen so that $m^2=0$ corresponds to the massless point, i.e. the point with the largest gauge symmetry.)
The shift symmetries of interest occur at the following special mass values,
\be \begin{cases} m_{[0],k}^2L^2=k(k+D-1),\ \ \ \ \ \ \ \ \ \ \ \ \ \ \ \ s=0\, , \\
m^2_{[s],k}L^2= (k+2) (k+D-3+2 s) ,\ \  s\geq 1\, ,
\end{cases} \ \ \  \ \ \ \ k=0,1,2,\ldots \,. \label{symtenpmmassee}
\ee
The shift symmetry reads
\be \label{eq:tensorshift}
\delta \phi_{\mu_1\cdots\mu_s}= S_{A_1\cdots A_{s+k},B_1\cdots B_s}X^{A_1}\cdots X^{A_{s+k}} {\partial X^{B_1}\over \partial x^{\mu_1}}\cdots {\partial X^{B_s}\over \partial x^{\mu_s}} \, ,
\ee
where the $X^A$ are embedding coordinates of AdS$_D$ into a flat spacetime of dimension $D+1$, and the tensor $S_{A_1\cdots A_{s+k},B_1\cdots B_s}$ is a fully traceless, constant embedding space tensor with the index symmetries of the following Young tableau,
\be 
S_{A_1\cdots A_{s+k},B_1\cdots B_s}\in ~\raisebox{1.15ex}{\gyoung(_5{s+k},_3s)} \, .
\ee
As explained in \cite{Bonifacio:2018zex}, the $k$-th shift symmetric spin $s$ field can be thought of as the missing longitudinal mode of a partially massless (PM) spin $s+k+1$ field of depth $s$ (i.e. it has a rank $s$ gauge parameter).  There is a symmetric traceless `field strength' with the same indices as this partially massless field, made from $k+1$ derivatives of the shift symmetric field in a way that mirrors the PM gauge transformation, 
\be F_{\mu_1\ldots \mu_{s+k+1}}= \nabla_{(\mu_{s+1}} \cdots \nabla_{\mu_{s+k+1}}\phi_{\mu_1\ldots \mu_s)_T}\,  . \ee
This field strength is invariant under the shift symmetries \eqref{eq:tensorshift} and gives the basic on-shell non-trivial shift-invariant operator in the theory.

Here, we will extend the results of \cite{Bonifacio:2018zex} by finding the analogs of the above shift symmetries for all the remaining bosonic fields, the anti-symmetric $p$-form fields and the mixed symmetry fields.  In the process, we will also see that there are no further shift symmetries of this type beyond those we find, and none further for the symmetric tensors beyond those in \eqref{symtenpmmassee}.

We will find these shift invariant fields as longitudinal modes of various mixed symmetry fields as they approach PM points. In what follows, we will therefore make frequent use of the classification of PM points for general mixed symmetry fields \cite{Metsaev:1995re,Metsaev:1997nj,Alkalaev:2003qv,Boulanger:2008up,Boulanger:2008kw,Skvortsov:2009zu,Skvortsov:2009nv,Basile:2016aen}.  A short summary is as follows.  Consider a general massive field with the symmetries of a $p$ row Young tableau with row lengths $[s_1,s_2,\ldots,s_p]$, which on-shell is traceless and divergenceless in all indices and is annihilated by the Klein-Gordon operator $\nabla^2-\tilde m^2$.  This field has dual conformal field theory (CFT) dimensions $\Delta_\pm$ found from the mass $\tilde m^2$ by finding the greater and lesser roots of
\be \tilde m^2L^2=\Delta(\Delta-d ) -\sum_{i=1}^p s_i . \label{mixsymmassderee}\ee
The partially massless points occur when squares in the tableau from a row which is longer than the row below it are `activated'.   If it is the $q$-th row that is being activated, then the number of squares that can be activated ranges from $1,2,\ldots, s_q-s_{q+1}$.  We assign a depth $t=0,1,\cdots, s_q-s_{q+1}-1$, which
indicates that $s_q-s_{q+1}-t$ of the squares in the $q$-th row are activated.   By removing the activated squares, we get the Young tableau of the gauge parameter, and each activated square becomes a derivative in the gauge transformation law.  
These PM points occur at integer values of the dual CFT dimension given by
\be \Delta_+= d - q + s_{q+1} + t\,.\ee
The gauge transformation is generally reducible, with $q-1$ levels of gauge-for-gauge parameters.  It will be important that the only irreducible gauge symmetries are for $q=1$, i.e. when the first row is activated; it is these whose longitudinal modes will give rise to the shift symmetric fields.

\vspace{.15cm}
\noindent
{\bf Conventions:}
The spacetime dimension is $D$, with indices $\mu,\nu,\ldots$.   The dual CFT dimension is $d \equiv D-1$, with indices $i,j,\ldots$. We use the mostly plus metric signature. 
We denote the AdS$_D$ radius by $L$, so that the Ricci scalar is $R=-{D(D-1)/ L^2}<0$.  Though we write everything in terms of AdS, our results also apply to dS with the replacement $L^2\rightarrow -1/H^2$ with $H$ the dS Hubble scale.  $X^A$ denotes the embedding of AdS$_D$ into flat spacetime of dimension $D+1$, with indices $A,B,\ldots$ (see appendix A of \cite{Bonifacio:2018zex} for details and conventions of the embedding formalism).  

Tensors are symmetrized and antisymmetrized with unit weight, {\it e.g.} $t_{[\mu\nu]}=\half \left(t_{\mu\nu}-t_{\nu\mu}\right)$, and $(\cdots)_T$ means the symmetric fully traceless part of the enclosed indices.  Young tableaux are denoted $[s_1,s_2,\ldots,s_p]$ where $s_i$ is the number of boxes in the $i$-th row, and are deployed in the manifestly anti-symmetric convention.  We sometimes use the shorthand of using an exponent to denote multiple rows of the same length, e.g. $[4,2^3,1]\equiv [4,2,2,2,1]$.
We denote the corresponding Young projectors by ${\cal Y}_{[s_1,s_2,\ldots,s_p]}$.  The notation $T$ on a tableau or projector indicates that it is fully traceless.  

Masses for mixed symmetry fields are denoted by $\tilde m^2$, which are the ``bare masses'' that appear in the Klein-Gordon equation satisfied by the transverse and traceless field: $\nabla^2-\tilde m^2=0$.  For symmetric tensors and $p$-forms, there is a traditional notion of ``massless,'' which is the partially massless point with the largest depth (the only partially massless point in the $p$-form case), so in these cases we use this more traditional $m^2$ which is shifted relative to $\tilde m^2$ so that $m^2=0$ corresponds to this massless point.

\section{Shift symmetries for 2-forms\label{2formsection}}

We start by illustrating the general pattern with the simplest case not covered by \cite{Bonifacio:2018zex}, the massive 2-form field.  We will do the simplest instance of this case fully off-shell at the Lagrangian level, then proceed to more on-shell methods as we go on to more general cases.

Consider the Lagrangian for a $2$-form field $B_{\mu\nu}$ of mass $m$ on AdS$_D$,
\be {1\over \sqrt{-g}}{\cal L}_{[1,1],m^2}(B)= -{1\over 12} (dB)_{\mu\nu\rho}^2-{1\over 4}m^2 B_{\mu\nu}^2\,,\ \ \ B_{\mu\nu}\in ~\raisebox{1.15ex}{\yng(1,1)} \,  ,\label{shiftsymmese}
\ee
where $(dB)_{\mu\nu\rho}=3\nabla_{[\mu} B_{\nu\rho]}$ is the field strength.  The equations of motion can be cast in the form
\be \left(\nabla^2 +{2(D-2) \over L^2}-m^2\right)B_{\mu\nu}=0,\ \ \ \nabla^{\nu} B_{\nu\mu}=0.\ee
The mass is defined such that when $m^2=0$ we get the usual massless $2$-form gauge symmetry.   There are no other points of enhanced gauge symmetry besides this.

As we will see, the massive 2-form theory \eqref{shiftsymmese} gets an enhanced shift symmetry when $m^2$ is set to the following values,
\be m_{[1,1],k}^2L^2=  \left(k+3 \right) \left(k+D-2\right)  ,\ \ \ k=0,1,2,\ldots\,.\label{2formmassvaluese}\ee
The form of the $k$-th shift symmetry is 
\be  \delta B_{\mu\nu}=  S_{B_1\ldots B_{k+1},A_1,A_2 }X^{B_1}\cdots X^{B_{k+1}} {\partial X^{A_1}\over \partial x^{\mu} } {\partial X^{A_2}\over \partial x^{\nu} } \,,\label{Bfieldshiftformee}  \ee
where $S_{B_1\ldots B_{k+1},A_1,A_2 }$ is a constant, fully traceless mixed symmetry embedding space tensor of type 
\be  
S_{B_1\ldots B_{k+1},A_1,A_2 }\in ~\raisebox{1.15ex}{\gyoung(_5{k+1},\ ,\ )} \, ,
\label{repKexpre}
\ee
and there is a $k+1$ derivative shift invariant field strength which is a traceless $[k+2,1]$ tensor 
\be F_{\mu\nu\rho_1\ldots\rho_{k+1}}={\cal Y}_{[k+2,1]}^T \left[\nabla_{\rho_1}\cdots\nabla_{\rho_{k+1}} B_{\mu\nu}\right] \in  ~\raisebox{1.15ex}{\gyoung(\mu  _5{k+1},\nu  )}\,. \label{21shiftfieldstrength3e}\ee

We will find these shift symmetries by considering partially massless limits of appropriate mixed symmetry fields.
For $k=0$ the appropriate field is a massive $[2,1]$ hook field (also known as a Curtright field \cite{Curtright:1980yk,Curtright:1980yj}).  The Lagrangian for a massive $[2,1]$ hook field on AdS$_D$ is \cite{Zinoviev:2002ye,Joung:2016naf},
\bea   {1\over \sqrt{-g}}{\cal L}_{[2,1],\tilde m^2}(f)&=& -{3\over 4} \left( \nabla_{[\mu} f_{\nu\rho]\sigma}\nabla^{[\mu} f^{\nu\rho]\sigma}-3  \nabla_{[\mu} f_{\nu\rho]}^{\ \ \ \rho}  \nabla^{[\mu} f^{\nu\sigma]}_{\ \ \ \ \sigma}\right) \nn\\
&&  -\frac{1}{4} \left({ 2 D-3 \over L^2} +\tilde m^2\right)\left(f_{\mu\nu\rho}f^{\mu\nu\rho}-2 f_{\mu\nu}^{\ \ \ \nu} f^{\mu\rho}_{\ \ \ \rho}\right)\,,\ \ \ f_{\mu\nu\rho}\in ~\raisebox{1.15ex}{\gyoung(\mu\rho,\nu)}\,. \nn\\ \label{hooklage}
\eea
The equations of motion can be cast in the form
\be \left(\nabla^2 -\tilde m^2\right)f_{\mu\nu\rho}=0\, ,\ \ \ f_{\mu\nu}^{\ \ \nu}=0\, ,\ \ \ \nabla^\mu f_{\mu\nu\rho}=\nabla^\rho f_{\mu\nu\rho}=0\, ,\ee
so the field on shell is fully traceless, fully divergenceless and satisfies a Klein Gordon equation with the bare mass $\tilde m^2$.  

The theory \eqref{hooklage} has two mass values at which partially massless gauge symmetries arise \cite{Brink:2000ag}:  
\begin{itemize}
\item
The first is where the top block is activated, giving an antisymmetric tensor gauge symmetry,  
\be ~\raisebox{1.15ex}{\gyoung(\ \nabla ,\ )}\ \ \ \tilde m^2L^2=-3\ :\ \ \ \delta f_{\mu\nu\rho}=\nabla_{[\mu}\Lambda_{\nu]\rho}-\nabla_\rho \Lambda_{\mu\nu},\ \ \ \Lambda_{\mu\nu}\in ~\raisebox{1.15ex}{\yng(1,1)} \, .\label{PMpoint2form1e}\ee
This gauge symmetry has no gauge-for-gauge reducibilities.  This mass value is unitary on AdS and non-unitary on dS.  
\item
The second is where the bottom block is activated, giving a symmetric tensor gauge symmetry, 
\be ~\raisebox{1.15ex}{\gyoung(\ \ ,\nabla )}\ \ \ \tilde m^2L^2=-(2D-3)\ :\ \ \ \delta f_{\mu\nu\rho}=\nabla_{[\mu}\xi_{\nu]\rho},\ \ \ \xi_{\mu\nu}\in ~\raisebox{0.0ex}{\yng(2)} \, .  \label{PMpoint2form1e22} \ee
This gauge symmetry has a gauge-for-gauge reducibility where
\be \delta \xi_{\mu\nu}=\nabla_\mu\nabla_\nu\chi-{1\over L^2}g_{\mu\nu}\chi\, ,\label{gaugeforgaugepme}\ee
with scalar gauge-for-gauge parameter $\chi$.  This mass value is unitary on dS and non-unitary on AdS.  
\end{itemize}

As we approach these partially massless points with the curvature scale $L$ held fixed, the longitudinal modes that are removed by the gauge symmetries must decouple \cite{Zinoviev:2001dt,Zinoviev:2002ye,Zinoviev:2003dd,Zinoviev:2008ze,Zinoviev:2008ve,Zinoviev:2009gh,DeRham:2018axr}.  Consider first the decoupling limit where we approach the first partially massless value \eqref{PMpoint2form1e} from above,
\be  \tilde m^2=-{3\over L^2}+\epsilon^2,\ \ \ \ \epsilon\rightarrow 0.\label{stukclimitee}\ee
To preserve the degrees of freedom in this limit we introduce a $2$-form St\"uckelberg field $B_{\mu\nu}$ and make the St\"uckelberg replacement
\be f_{\mu\nu\rho}\rightarrow   f_{\mu\nu\rho}+{1\over \epsilon}\left( \nabla_{[\mu}B_{\nu]\rho}-\nabla_\rho B_{\mu\nu}\right),\ \ \ B_{\mu\nu}\in ~\raisebox{1.15ex}{\yng(1,1)} \, ,\ee
where we have inserted the factor of $1/\epsilon$ so that $B_{\mu\nu}$ will come out canonically normalized up to numerical factors.
This replacement introduces a St\"uckelberg gauge symmetry under which the St\"uckelberg field shifts,
\be \delta  f_{\mu\nu\rho}=\left( \nabla_{[\mu}\Lambda_{\nu]\rho}-\nabla_\rho \Lambda_{\mu\nu}\right),\ \ \ \delta B_{\mu\nu}=-\epsilon\,\Lambda_{\mu\nu}.\ee
In the limit \eqref{stukclimitee}, the Lagrangian \eqref{hooklage} splits up into a partially massless hook \eqref{PMpoint2form1e} and a correct-sign massive 2-form with mass $m^2L^2= 3\left(D-2\right) $,
\be{ \cal L}_{[2,1],\tilde m^2}(f) \underset{\tilde m^2\rightarrow -{3\over L^2}}{\rightarrow} {\cal L}_{[2,1], -{3\over L^2}}(f) +{3\over 2}\,{\cal L}_{[1,1], {3\left(D-2\right)\over L^2} }(B).\label{lagpm21splitwe}\ee
This mass is precisely the $k=0$ value of \eqref{2formmassvaluese}.

As discussed in \cite{Bonifacio:2018zex}, the shift symmetry arises from reducibility parameters of the St\"uckelberg gauge symmetry, i.e. values of $\Lambda_{\mu\nu}$ for which the gauge transformation of the hook field vanishes, $\nabla_{[\mu}\Lambda_{\nu]\rho}-\nabla_\rho \Lambda_{\mu\nu}=0.$  In terms of the embedding space field $\Lambda_{AB}$ corresponding to the gauge parameter $\Lambda_{\mu\nu}$, this equation says that the mixed symmetry part of $\partial_A \Lambda_{BC}$ should vanish.  In addition, $\Lambda_{AB}$ should be transverse to $X^A$, should satisfy the higher dimensional Klein-Gordon equation $\square_{(D+1)}=0$, and should be of weight $1$ in $X^A$ (so that the Klein-Gordon equation pulls back to the correct mass).  The solution for all this is $\Lambda_{A_1A_2}=S_{B_1A_1A_2 }X^{B_1}$ with totally antisymmetric $S_{B_1A_1A_2 }$, which when pulled back to AdS gives the $k=0$ case of \eqref{Bfieldshiftformee}\footnote{This is also known as a rank 2 Killing-Yano tensor and appears in generalized symmetries for linearized gravity \cite{Penrose:1986ca,Jezierski:2002mn,Kastor:2004jk,Jezierski:2014gka,Jezierski:2019xxl,Benedetti:2021lxj,Benedetti:2022zbb,Hinterbichler:2022agn}.} 
\be \Lambda_{\mu\nu}=S_{B_1A_1A_2}X^{B_1}{\partial X^{A_1}\over \partial x^{\mu} } {\partial X^{A_2}\over \partial x^{\nu} } .\label{21antsymreduee}\ee

This implies that there is a one-derivative `field strength' with the same symmetries as the parent PM field $f_{\mu\nu\rho}$, 
\be F_{\mu\nu\rho}\equiv \nabla_{[\mu}B_{\nu]\rho}-\nabla_\rho B_{\mu\nu}\,,\ee  
which is invariant under the $k=0$ shift symmetry \eqref{Bfieldshiftformee}.  The trace of this field strength is proportional to $\nabla^\nu B_{\mu\nu}$ and so vanishes on-shell.  The traceless part is the basic on-shell non-trivial shift invariant operator in the theory, and is the $k=0$ case of \eqref{21shiftfieldstrength3e}.

We can also consider a decoupling limit where we approach the second partially massless point \eqref{PMpoint2form1e22},
\be  \tilde m^2=-{2D-3\over L^2}+\epsilon^2,\ \ \ \ \epsilon\rightarrow 0.\label{stukclimitee2}\ee
In this case we introduce a symmetric St\"uckelberg field
\be f_{\mu\nu\rho}\rightarrow   f_{\mu\nu\rho}+{1\over \epsilon} \nabla_{[\mu}H_{\nu]\rho},\ \ \ H_{\mu\nu}\in ~\raisebox{0.0ex}{\yng(2)} \, .\ee
In the limit \eqref{stukclimitee2}, the Lagrangian \eqref{hooklage} splits up into a partially massless hook and a massive spin-2,
\be{ \cal L}_{[2,1],\tilde m^2}(f) \underset{\tilde m^2\rightarrow  -{2D-3\over L^2}}{\rightarrow} {\cal L}_{[2,1], -{2D-3\over L^2}}(f) +{1\over 4}\,{\cal L}_{[2], -{D-2\over L^2}}(H)\,,\ee
where ${\cal L}_{[2], m^2}$ is the Fierz-Pauli action for a massive spin-2 field on AdS$_D$ (as written in e.g. (5.2) of \cite{Hinterbichler:2011tt}).   This massive spin-2 that we get is not a new shift-symmetric field, rather its mass $m^2L^2=-(D-2)$ is that of the partially massless graviton \cite{Deser:1983tm,deRham:2013wv}.  This is because of the gauge-for-gauge reducibility \eqref{gaugeforgaugepme}, which shows up in the decoupling limit as a partially massless gauge symmetry for the longitudinal field $H_{\mu\nu}$.  

This illustrates a general point: the shift symmetric fields can come only from the missing longitudinal modes of PM fields with irreducible gauge symmetries, i.e. those in which the first row is activated.  PM field with gauge-for-gauge reducibilities, i.e. those where a row below the first is activated, instead spin off longitudinal modes which are themselves PM fields, and thus do not give new shift fields.  The gauge-for-gauge parameter $\chi$ in \eqref{gaugeforgaugepme} represents the longitudinal mode of the PM spin-2 field, so this is the $k=1$ shift symmetric scalar.  This illustrates another general point: the endpoint of gauge-for-gauge reducibility chains of partially massless fields is always a shift symmetric field.  But these do not give new shift symmetric fields because they are already accounted for by the longitudinal modes of irreducible PM fields.

We can see all of the above from the dual CFT perspective \cite{Alkalaev:2012ic,Alkalaev:2012rg}.  The massive $[2,1]$ field has a dual $[2,1]$ traceless primary state 
\be |f_{ijk}\rangle_\Delta \in ~\raisebox{1.15ex}{\gyoung(ik,j)} \, , \label{21operatorge}\ee 
where the mass and conformal scaling dimension are related by
\be \tilde m^2L^2=\Delta(\Delta-d ) -3\, . \ee
Denoting the larger and smaller roots of this as $\Delta_\pm$, the PM point of interest \eqref{PMpoint2form1e} gives 
\be \Delta_+=d\, .\ee
At this value, a $[2,1]$ state of type \eqref{21operatorge} saturates its unitary bound and develops null states in its Verma module, leading to a shortening condition.   In general, the level at which this shortening occurs is equal to the number of derivative in the PM gauge transformation law.  Since the gauge symmetry \eqref{PMpoint2form1e} has one derivative, the CFT state gets a conservation-type shortening condition at level one in the Verma module: $P^k|f_{ijk}\rangle_d=0$.  This is a null state of spin $[1,1]$ and dimension $d+1$ which spans its own sub-module.  As the PM value is approached, the AdS$_D$ representation $(\Delta,[2,1])$ spanned by the primary $|f_{ijk}\rangle_\Delta$ and its descendants splits according to the branching rule
\be (\Delta,[2,1])\underset{\Delta\rightarrow d}{\rightarrow} (d,[2,1]) \oplus (d+1,[1,1])\,. \label{gt21gt11e}\ee
The submodule spanned by the null states is $(d+1,[1,1])$.  

The mass of a 2-form is related to its conformal dimension by 
\be m^2L^2=(\Delta-2)(\Delta-d+2)\, ,\label{2formconfdimer}\ee
 and using this we see that the representation $(d+1,[1,1])$ is precisely the $\Delta_+$ value of a $k=0$ shift symmetric 2-form with mass as written in \eqref{2formmassvaluese}.  The expression \eqref{gt21gt11e} is the group theoretical version of the Lagrangian expression \eqref{lagpm21splitwe}.

If we consider the lesser root $\Delta_-=-1$ for the shift field, we get the non-unitary representation $(-1,[1,1])$, spanned by the primary $|b_{ij}\rangle_{-1}$.  This representation is finite dimensional once the null states are factored out; the only non-null states are
\be |b_{ij}\rangle_{-1} \in ~\raisebox{1.15ex}{\yng(1,1)}\, ,\ \  P^j|b_{ij}\rangle_{-1} \in ~\raisebox{0.0ex}{\yng(1)}\, ,\ \  P_{[k}|b_{ij]}\rangle_{-1} \in ~\raisebox{2.2ex}{\yng(1,1,1)}\, ,\ \ P_{[i}P^k|b_{j]k}\rangle_{-1} +{d-4\over d-1} P^2|b_{ij}\rangle_{-1} \in ~\raisebox{1.15ex}{\yng(1,1)}\, .\ee
These states join together into a $[1,1,1]$ in $d+2$ dimensions, so this is the finite dimensional $[1,1,1]$ representation of the AdS$_D$ isometry algebra $so(2,D-1)$, precisely the anti-symmetric tensor in \eqref{21antsymreduee} which parametrizes the shift symmetries.

To get the higher values of $k$ for the shift-symmetric 2 form, we start with a massive $[k+2,1]$ tableau field, 
\be f_{\mu\nu\rho_1\ldots\rho_{k+1}}\in  ~\raisebox{1.15ex}{\gyoung(\mu  _5{k+1},\nu  )},\ \ \ \left(\nabla^2-\tilde m^2\right) f_{\mu\nu\rho_1\ldots\rho_{k+1}}=0\, ,\label{k1ew2forme}\ee
which is also fully traceless and fully divergenceless on shell.  We consider the partially massless point where all the possible top blocks are activated, so that the gauge parameter is a 2-form,
\be \overset{\ \ \ \ \overbrace{\ \ \ \ \ \ \ \  \ \ \ \ \ \ \ \ }^{k+1} }{~\raisebox{1.15ex}{\gyoung(\  \nabla\nabla_2\cdots\nabla,\  )}} \ \  \tilde m^2=-{k+3\over L^2} \,:\ \ \ \delta  f_{\mu\nu\rho_1\ldots\rho_k}={\cal Y}_{[k+2,1]}^T \left[\nabla_{\rho_1}\cdots\nabla_{\rho_{k+1}} \Lambda_{\mu\nu}\right],\ \ \ \Lambda_{\mu\nu}\in ~\raisebox{1.15ex}{\yng(1,1)} \,. \ \label{genbfiledcopmee}\ee
This gauge symmetry is irreducible, so the resulting longitudinal mode in the partially massless limit will be a shift field rather than a gauge field.  

Using reasoning similar to the $k=0$ case we can find the reducibility parameters for the gauge symmetries \eqref{genbfiledcopmee} and they are the shifts \eqref{Bfieldshiftformee}.   There is a $k+1$ derivative field strength with the same symmetries as the PM field $f_{\mu\nu\rho_1\ldots\rho_{k+1}}$, made out of $k+1$ derivatives of the shift field patterned after the gauge transformation \eqref{genbfiledcopmee}, which is invariant under these shifts and gives \eqref{21shiftfieldstrength3e}.

A massive $[k+2,1]$ field has a dual $[k+2,1]$ primary state 
\be |f_{ij\, l_1\ldots l_{k+1} }\rangle_\Delta\, \in  ~\raisebox{1.15ex}{\gyoung(i_5{k+1},j)} ,\ee
where the mass and conformal scaling dimension are related by
\be \tilde m^2L^2=\Delta(\Delta-d ) -k-3. \ee
At the PM point of interest \eqref{genbfiledcopmee}, the conformal dimension is 
\be \Delta_+=d\, .\label{dualdvaluee11}\ee  
Since the gauge transformation \eqref{genbfiledcopmee} has $k+1$ derivatives, the dual state at the value \eqref{dualdvaluee11} is a kind of multiply-conserved current \cite{Brust:2016gjy} which gets a shortening condition at level $k+1$ in the Verma module,  $P^{l_1}\cdots P^{l_{k+1}}|f_{ij\, l_1\ldots l_{k+1} }\rangle_\Delta=0$.  This is a null state of spin $[1,1]$ and dimension $d+k+1$ which spans its own sub-module.  As the PM value is approached, the AdS$_D$ representation $(\Delta,[k+2,1])$ spanned by the primary $|f_{i_1i_2j_1\ldots j_{k+1} }\rangle_\Delta$ and its descendants splits according to the branching rule
\be (\Delta,[k+2,1])\underset{\Delta\rightarrow d}{\rightarrow} (d,[k+2,1]) \oplus (d+k+1,[1,1])\,. \ee
Using \eqref{2formconfdimer}, we see that the representation $(d+k+1,[1,1])$ is precisely the $\Delta_+$ value of a level $k$ shift symmetric 2-form as written in \eqref{2formmassvaluese}.  The negative root $\Delta_-$ gives the representation $(-k-1,[1,1])$ whose non-null states span the finite dimensional, non-unitary representation $[k+1,1,1]$ of the AdS$_D$ isometry group $so(2,D-1)$, and these are precisely the shift symmetry parameters \eqref{repKexpre}.

The PM point where the bottom block is activated,
\be  ~\raisebox{1.15ex}{\gyoung(\  _5{k+1},\Delta)}\ee
has a gauge-for-gauge reducibility, and the longitudinal mode will be a depth $t=0$ partially massless spin $k+2$ field, so this gives no new shift symmetric fields.

The PM points where fewer than the maximal number of top blocks are activated will have gauge parameters which are again mixed symmetry tensors, and since these gauge transformations are irreducible, these will give rise to shift-symmetric points for these mixed symmetry tensors.  We will return to this more general case in section \ref{gensection}, after discussing the higher $p$-forms in the next section.

We can ask if there are other possible shift-symmetric mass values for the 2-form besides those in \eqref{2formmassvaluese}, perhaps coming from PM limits of more complicated mixed symmetry tensors.  The answer is no for the following reason.   As we have seen, to get a shift-symmetric field for the longitudinal mode the PM gauge symmetry must be irreducible, otherwise the gauge-for-gauge reducibility parameters will become gauge symmetries of the longitudinal mode and we will get other PM fields rather than shift fields.  The depth of gauge-for-gauge redundancy in a PM field is equal to the row number which is activated \cite{Skvortsov:2009zu}.  The PM symmetry must therefore come from a PM tableau where only the top row is activated, so that the gauge symmetry is irreducible.  To get a shift-symmetric 2-form, we need a PM tableau whose first row is activated and whose gauge parameter is a 2-form.  The only such tableaux are those in \eqref{genbfiledcopmee}.  The same reasoning shows that the shift symmetric points \eqref{symtenpmmassee} for symmetric tensor fields, and the shift symmetric points for more general fields that we find below, are the only ones.

\section{Shift symmetries for $p$-forms}

We now move on to a massive $p$-form field $B_{\mu_1\ldots \mu_p}$ on AdS$_D$, $p\geq 1$.  The equations of motion can be cast in the form
\be \left(\nabla^2 +{1\over L^2}p(D-p)-m^2\right)B_{\mu_1\ldots \mu_p}=0,\ \ \ \nabla^{\mu_1} B_{\mu_1\ldots \mu_p}=0,\ \   \ B_{\mu_1\ldots \mu_p} \in  ~\raisebox{2.15ex}{\gyoung(|3p)}\ \ .\label{massivepformaee}\ee
The mass is defined such that at $m^2=0$ we get the usual massless $p$-form gauge symmetry.  There are no other points of enhanced gauge symmetry besides this.

The massive $p$-forms \eqref{massivepformaee} get an enhanced shift symmetry at the following mass values
\be m_{[1^p],k}^2L^2= \left( k + p+1\right) \left( k - p+D\right) ,\ \ \ k=0,1,2,\ldots\, ,\label{pformmassvaluese}\ee
and the form of the shift symmetry is 
\be  \delta B_{\mu_1\ldots \mu_p}=  S_{B_1\ldots B_{k+1},A_1,\ldots ,A_p } X^{B_1}\cdots X^{B_{k+1}}  {\partial X^{A_1}\over \partial x^{\mu_1} } \cdots {\partial X^{A_p}\over \partial x^{\mu_p} }\,,\label{Bfieldshiftfpormee}  \ee
where $S_{B_1\ldots B_{k+1},A_1,\ldots ,A_p }$ is a constant fully traceless mixed symmetry embedding space tensor of type $[k+1,1^p]$,
\be  
S_{ B_1\ldots B_{k+1}, A_1,\ldots ,A_p}\in ~\raisebox{2.15ex}{\gyoung(_5{k+1},|3p)} \, .
\label{repKexprepee}
\ee
There is an invariant field strength of type $[k+2,1^{p-1}]$,
\be F_{\nu_1\ldots \nu_{k+1}\mu_1, \ldots ,\mu_p }={\cal Y}_{[k+2,1^{p-1}]}^T\left[ \nabla_{\nu_1}\cdots\nabla_{\nu_{k+1}} B_{\mu_1\ldots \mu_p }\right] \in ~\raisebox{2.15ex}{\gyoung(|4p_4{k+1}) } \,. \ee

These shift fields come from the longitudinal modes of mixed symmetry PM fields of the form $[k+2,1^{p-1}]$,
\be  
f_{ \nu_1\ldots \nu_{k+1}\mu_1,\ldots ,\mu_p}\in ~\raisebox{2.15ex}{\gyoung(|4p_4{k+1}) } \, ,\ \   \left(\nabla^2-\tilde m^2\right) f_{ \nu_1\ldots \nu_{k+1}\mu_1,\ldots, \mu_p}=0\,,
\ee
in which all the possible blocks in the upper row are activated,
\bea &&\overset{\ \ \ \overbrace{\ \ \ \ \ \ \ \ \ \ \ \ \ \ \ \ }^{k+1} }{~\raisebox{2.15ex}{\gyoung(|4p\nabla\nabla_2\cdots\nabla) }}\,\ \   \tilde m^2_{\rm PM}=-{p+k+1\over L^2}
\,: \ \ \ \nn\\ 
&&\delta f_{ \nu_1\ldots \nu_{k+1}\mu_1,\ldots, \mu_p}={\cal Y}_{[k+2,1^{p-1}]}^T\left[ \nabla_{\nu_1}\cdots\nabla_{\nu_{k+1}} \Lambda_{\mu_1\ldots \mu_p }\right] ,\ \ \ \Lambda_{\mu_1\ldots \mu_p} \in  ~\raisebox{2.15ex}{\gyoung(|3p)}\, . \label{genbfiledcopmep2e}\eea
The shift symmetries \eqref{Bfieldshiftfpormee} are the reducibility parameters of this PM transformation. 

A massive $[k+2,1^{p-1}]$ field has a dual primary operator with conformal scaling dimension related to the mass by
\be \tilde m^2L^2=\Delta(\Delta-d ) -(k+p+1)  \, . \ee
At the PM point of interest \eqref{genbfiledcopmep2e}, the conformal dimension is 
\be \Delta_+=d\, .\ee  
At this value, since the gauge symmetry has $k+1$ derivatives the dual operator gets a conservation-type shortening condition at level $k+1$, giving a null state of spin $[1^p]$ and dimension 
\be \Delta_+=d+k+1.\ee  
As the PM value \eqref{genbfiledcopmep2e} is approached, the AdS$_D$ representation $(\Delta,[k+2,1^{p-1}])$ splits according to the branching rule
\be (\Delta,[k+2,1^{p-1}])\underset{\Delta\rightarrow d}{\rightarrow} (d,[k+2,1^{p-1}]) \oplus (d+k+1,[1^p]])\,. \ee
The relation between the mass and dual CFT scaling dimension of a $p$-form field is given by
\be m^2L^2=(\Delta-p)(\Delta-d+p) .\ee
Using this, we see that the representation $(d+k+1,[1^p])$ is precisely the $\Delta_+$ value of a level $k$ shift symmetric $p$-form with mass value as written in \eqref{pformmassvaluese}.  These values are all above the unitarity bound $\Delta\geq d-p$ for a $p$-form, indicating that the shift fields are unitary on AdS, and irreducible with no further null states.  The negative root $\Delta_-$ gives the representation $(-k-1,[1^p])$ whose non-null states span the finite dimensional, non-unitary representation $[k+1,1^p]$ of the AdS$_D$ isometry group $so(2,D-1)$, and these are precisely the shift symmetry parameters \eqref{repKexprepee}.

\begin{figure}
\begin{center}
\epsfig{file=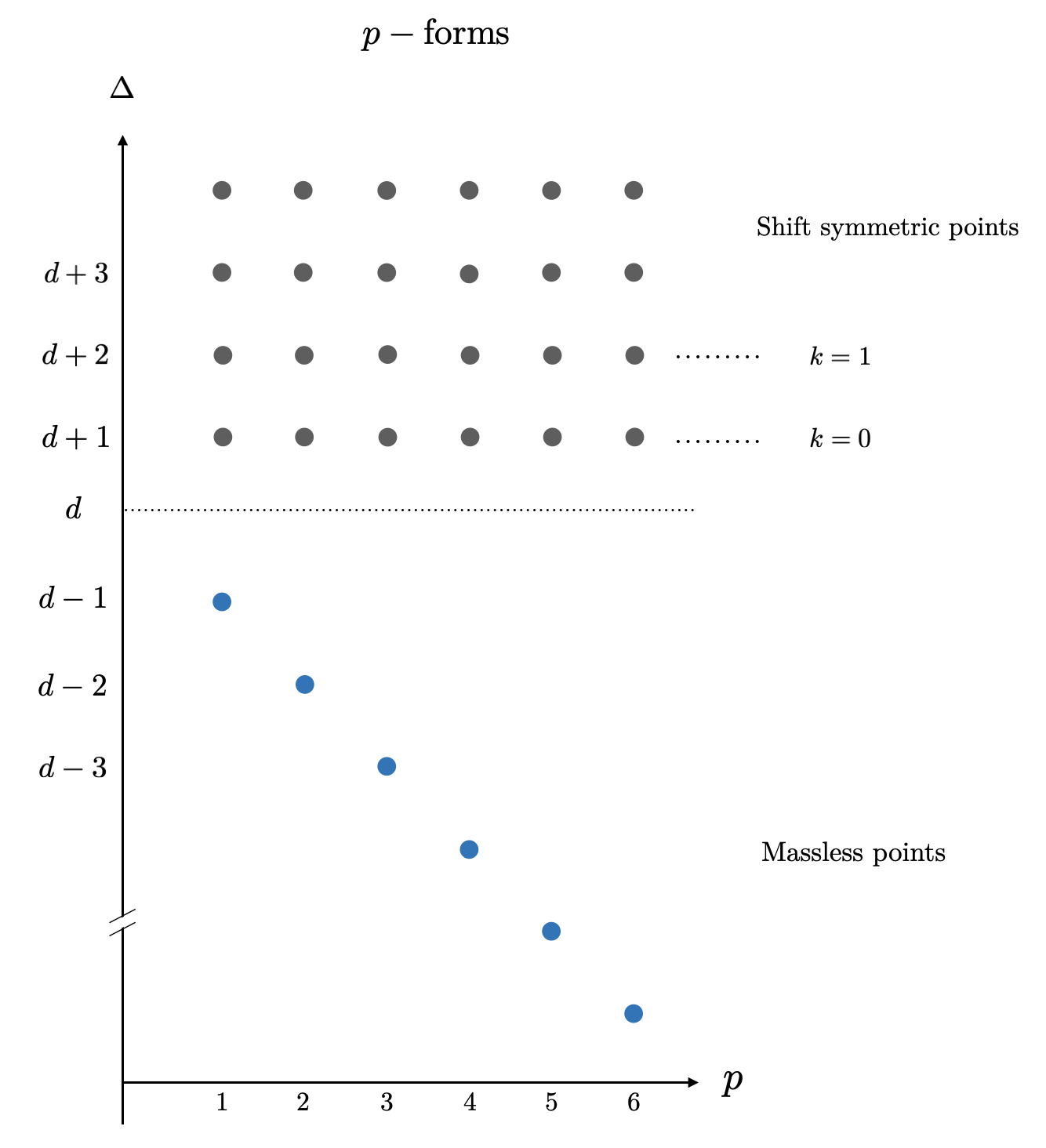,width=5in}
\caption{\small $p$ form fields in the conformal dimension $\Delta$ vs. $p$ plane, for the values $\Delta_+$.  The masses are given by $m^2L^2=(\Delta-p)(\Delta-d+p)$.  Blue dots are the massless points, black dots are shift symmetric points.  For each $p$ the unitarity bound coincides with the massless point.}
\label{pformfigure}
\end{center}
\end{figure}

The shift symmetric $p$-form values are summarized in figure \ref{pformfigure}.

\section{Shift symmetries for general mixed symmetry fields\label{gensection}}

We now turn to the general mixed symmetry case. Consider a general massive field in a tableau $[s_1,s_2,\ldots,s_p]$ which satisfies on-shell the Klein-Gordon equation,
\be \phi_{\mu_1\ldots\mu_{s_1},\ldots}\in  \raisebox{-40pt}{\epsfig{file=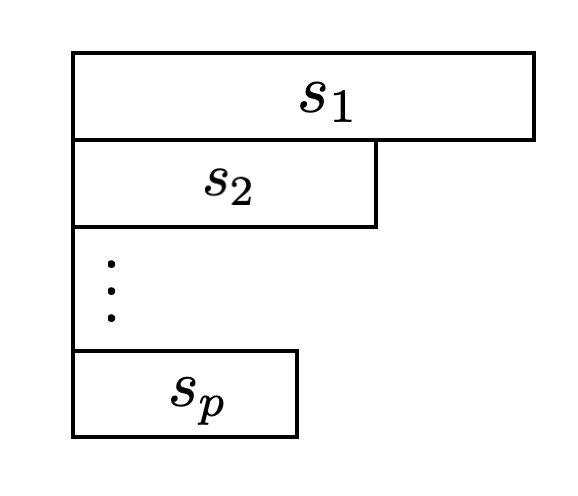,height=1.2in,width=1.2in}}  \label{genshifttypee}\, ,\ \ \ \  \left(\nabla^2-\tilde m^2\right)  \phi_{\mu_1\ldots\mu_{s_1},\ldots}=0\, ,\ee  
and in addition is also fully traceless and fully divergenceless.  This field has a dual conformal dimension related to the mass $\tilde m^2$ by 
\be \tilde m^2L^2=\Delta(\Delta-d ) -\sum_{i=1}^p s_i . \label{mixsymmassderee2}\ee

We find the shift symmetry values by considering the PM fields of type $[s_1+k+1,s_2,\ldots,s_p]$ where $k+1$ boxes in the top row are activated (depth $t=s_1-s_2$),
\be  \raisebox{-40pt}{\epsfig{file=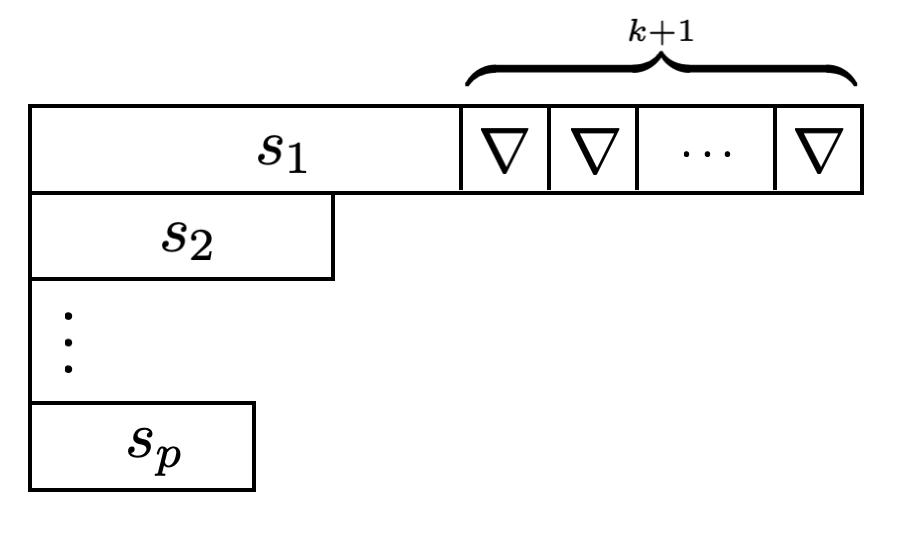,height=1.4in,width=2.3in}} ,\ \ k=0,1,2,\ldots\, ,   \ee 
since these are the only PM fields with irreducible gauge symmetries whose gauge parameter is of type $[s_1,s_2,\ldots,s_p]$.  These partially massless points occur at the mass values
\be \tilde m^2_{\rm PM}L^2= (s_1+D-2) (s_1-1)  - k-1 - \sum_{i=1}^ps_i\, .\ee
At these values, the dual CFT conformal dimension is 
\be \Delta_+=d - 1 +  s_1\,.\ee
Since the gauge symmetry has $k+1$ derivatives the dual operator gets a conservation-type shortening condition at level $k+1$, giving a null state of spin $[s_1,s_2,\ldots, s_p]$ and dimension 
\be \Delta_+=d +k  +  s_1.\label{genmixwsconfde}\ee  
As the PM value is approached, the AdS$_D$ representation $(\Delta,[s_1+k+1,s_2,\ldots, s_p])$ splits according to the branching rule
\be (\Delta,[s_1+k+1,s_2,\ldots, s_p])\underset{\Delta\rightarrow d - 1 +  s_1 }{\rightarrow} ( d - 1 +  s_1 ,[s_1+k+1,s_2,\ldots, s_p]) \oplus (d +k  +  s_1,[s_1,s_2,\ldots, s_p])\,. \ee
The representation $(d +k  +  s_1,[s_1,s_2,\ldots, s_p])$ is the $\Delta_+$ value of a level $k$ shift symmetric field corresponding to the longitudinal mode.  Using \eqref{mixsymmassderee2} we can translate this into the masses for the shift symmetric fields of type $[s_1,\ldots, s_p]$, 
\be \tilde m^2_{[s_1,\ldots, s_p],k}L^2=(s_1+k+D-1)(s_1+k)-  \sum_{i=1}^ps_i \,. \ \label{gensymshsmassee}\ee

The form of the shift symmetry is given by a constant fully traceless embedding space tensor of type $[s_1+k,s_1,s_2,\ldots,s_p]$, where the indices of the top row are contracted with $X^A$ and the rest are projected down,
\be \delta\phi_{\mu_1\ldots\mu_{s_1},\ldots}=S_{A_{1}\ldots A_{{s_1+k}},  B_{1}\ldots B_{{s_1}},\ldots} X^{A_{1}}\cdots X^{A_{s_1+k}}{\partial X^{B_{1}}\over \partial x^{\mu_1}}\cdots {\partial X^{B_{s_1}}\over \partial x^{\mu_{s_1}}}\cdots \, \ ,\label{embeddingbpgeneralpe}
\ee
\be S_{A_{1}\ldots A_{\mu_{s_1+k}},  B_{\mu_1}\ldots B_{\mu_{s_1}},\ldots} \in  \raisebox{-50pt}{\epsfig{file=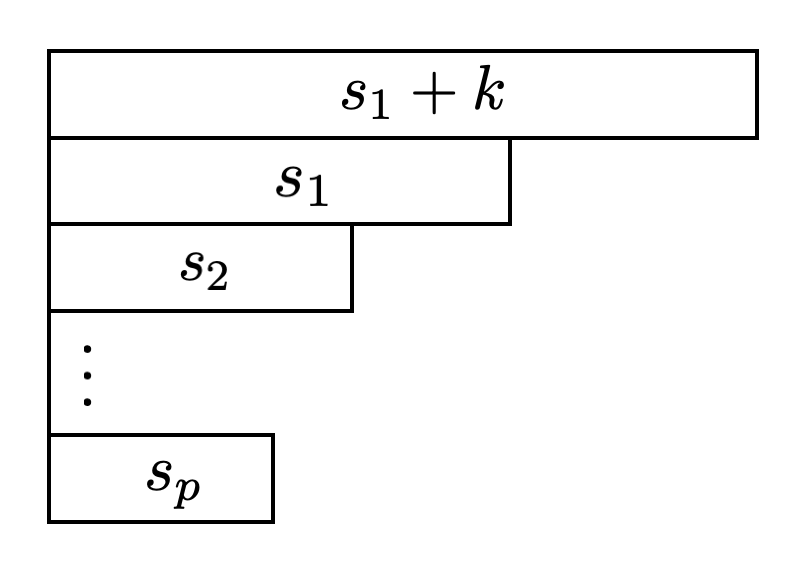,height=1.4in,width=1.6in}} \, \ .\label{mixedsymshifspe}\ee 
 The embedding space tensor $S_{A_{1}\ldots A_{{s_1+k}},  B_{1}\ldots B_{{s_1}},\ldots} X^{A_{1}}\ldots X^{A_{s_1+k}}$ that projects to \eqref{embeddingbpgeneralpe} is transverse to $X^A$ due to the mixed symmetry of the coefficients \eqref{mixedsymshifspe}, satisfies the ambient massless Klein-Gordon equation $\square_{(D+1)}=0$ due to the tracelessness of the coefficients \eqref{mixedsymshifspe}, and has homogeneity degree $w=s_1+k$ in the $X^A$.  The ambient Klein-Gordon operator acting on a degree $w$ tensor of type $[s_1,\ldots,s_p]$ reduces to the AdS$_D$ surface as $\square_{(D+1)} \rightarrow \nabla^2-{1\over { L}^2}\left(w(D+w-1)-\sum_{i=1}^p s_i \right)$.  Using $w=s_1+k$ this reproduces the masses \eqref{gensymshsmassee}.  We can think of the right hand side of \eqref{embeddingbpgeneralpe} as the most general kind of Killing-Yano-like object, or spherical harmonic, on AdS$_D$.
 
  The negative root $\Delta_-=-k  -  s_1$ of the shift field gives the representation $(-k  -  s_1,[s_1,s_2,\ldots, s_p])$ whose non-null states span the finite dimensional, non-unitary representation $[s_1+k,s_1,s_2,\ldots, s_p]$ of the AdS$_D$ isometry group $so(2,D-1)$, and these are precisely the shift symmetry parameters \eqref{mixedsymshifspe}.

 There is a $k+1$ derivative `field strength' with the symmetries $[s_1+k+1,s_2,\cdots,s_p]$ of the parent PM field which is invariant under the shifts \eqref{embeddingbpgeneralpe},
 \be F_{\mu_1\cdots\mu_{s_1+k+1},\cdots} = {\cal Y}_{[s_1+k+1,s_2,\cdots,s_p]}^T \left[ \nabla_{\mu_{s_1+1}}\cdots\nabla_{\mu_{s_1+k+1}} \phi_{\mu_1\cdots\mu_{s_1},\cdots}  \right] \,.\label{genmixsyfise}\ee
This is the basic local operator which captures the on-shell non-trivial shift invariant information in the theory.

\begin{figure}
\begin{center}
\epsfig{file=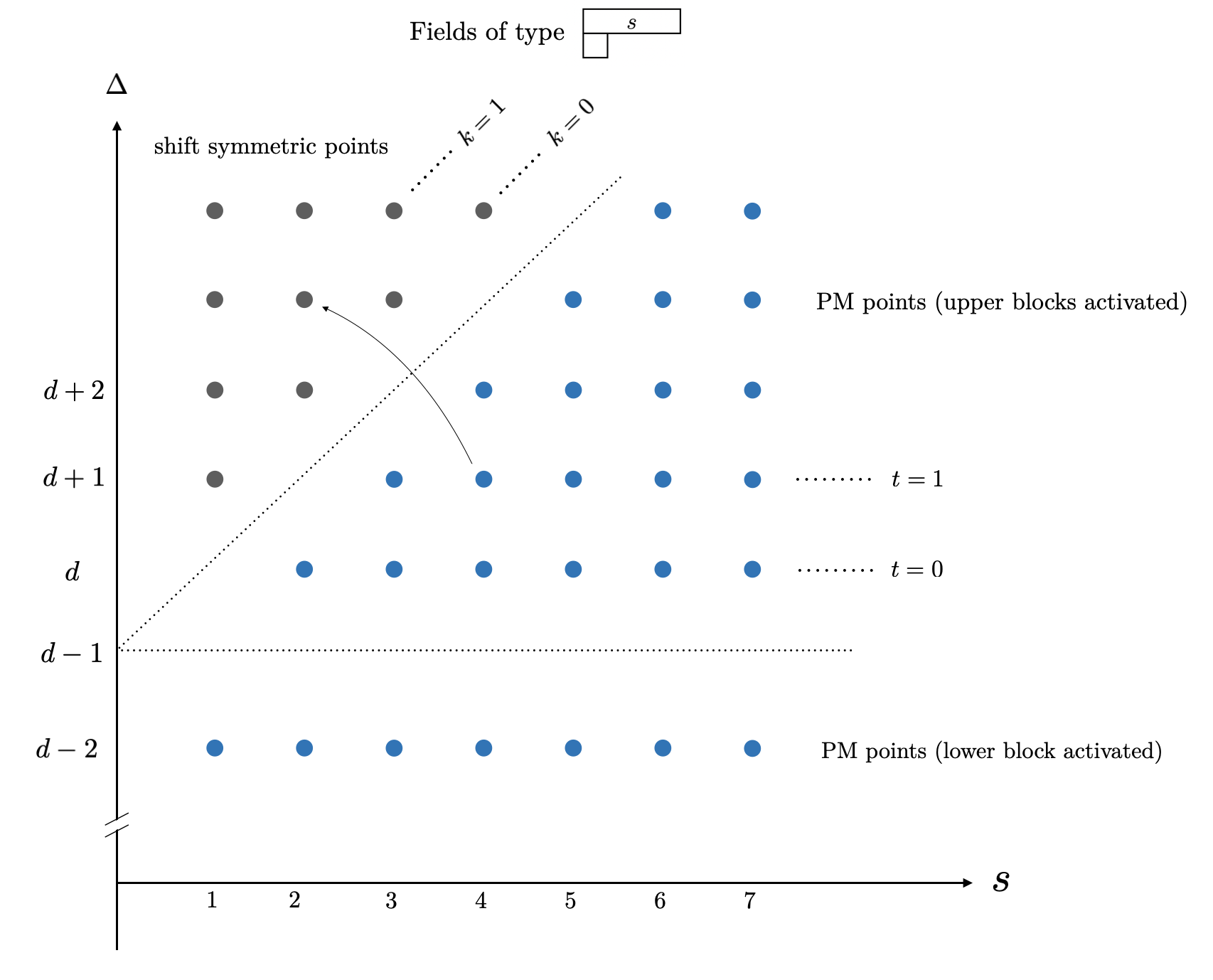,width=6in}
\caption{\small Fields of symmetry type $[s,1]$ in the conformal dimension $\Delta$ vs. $s$ plane, for the values of the positive root $\Delta_+$.  The masses are given by $\tilde m^2L^2=\Delta(\Delta-d ) -s-1$.
Blue dots are PM points, black dots are shift symmetric points.  The shift symmetric points are the longitudinal modes of the PM points where the upper row is activated.  The shift point field corresponding to a given such PM point is found by reflecting about the line $\Delta=s+d-1$, as illustrated by the curved arrow.  The AdS unitarity bound for each $s$ coincides with the uppermost PM point.}
\label{s1figure}
\end{center}
\end{figure}

\begin{figure}
\begin{center}
\epsfig{file=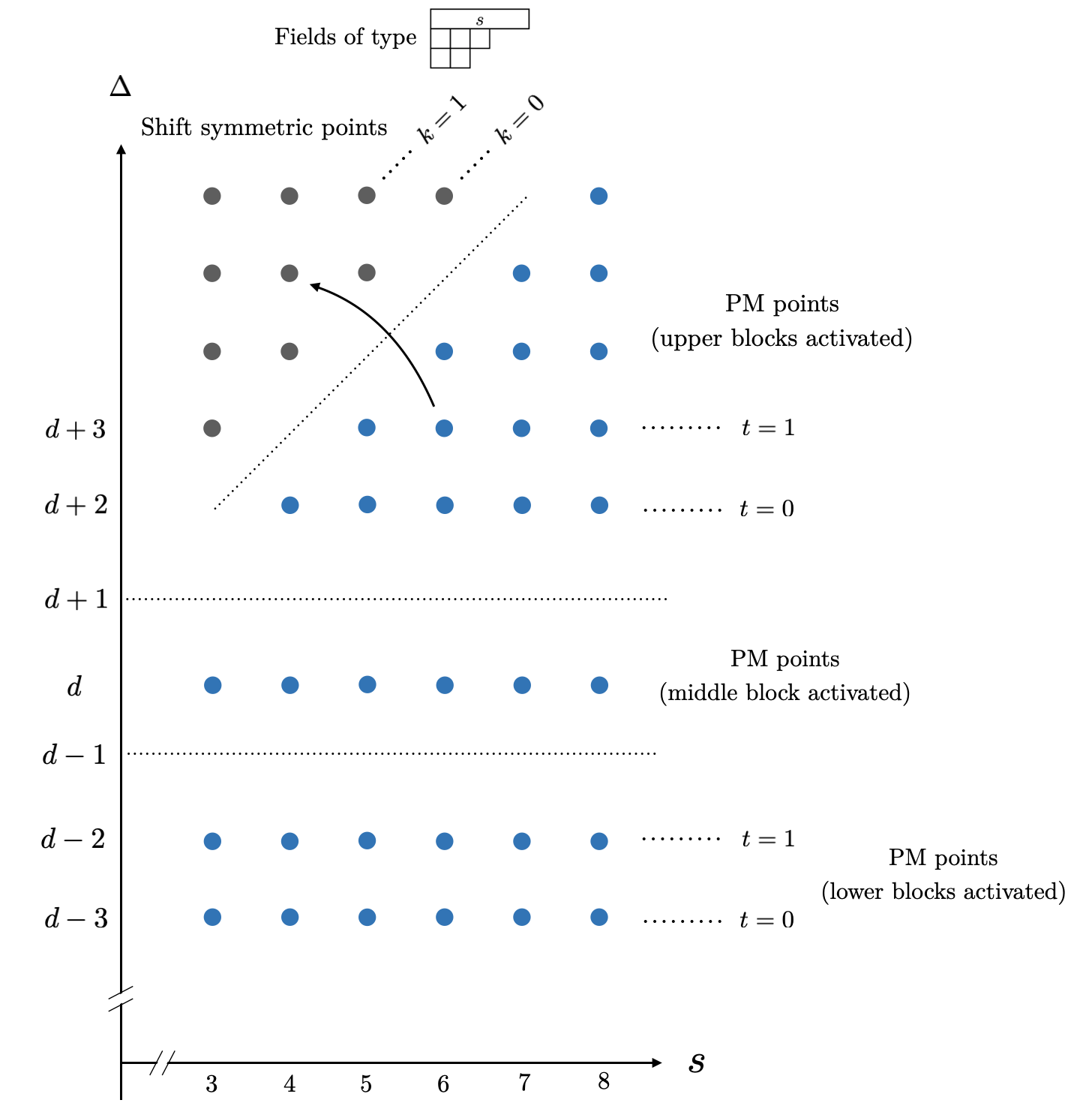,width=6in}
\caption{\small Fields of symmetry type $[s,3,2]$ in the conformal dimension $\Delta$ vs. $s$ plane, for the values $\Delta_+$.  The masses are given by $\tilde m^2L^2=\Delta(\Delta-d ) -s-5$.  Blue dots are PM points, black dots are shift symmetric points.  The shift symmetric points are the longitudinal modes of the PM points where the upper row is activated.  The shift point corresponding to a given such PM point is found by reflecting about the line $\Delta=s+d-1$, as illustrated by the curved arrow.  The AdS unitarity bound for each $s$ coincides with the uppermost PM point.}
\label{s32figure}
\end{center}
\end{figure}

As an illustrative example, let us return to the PM $[s_1,1]$ fields studied in Section \ref{2formsection}, and consider the PM points not used there, the ones where not all of the top blocks are activated.
The $t=s-1$ first row partially massless point of a $[s+k+1,1]$ field,
\be \overset{\ \ \ \ \ \ \ \ \ \ \ \ \ \ \overbrace{\ \ \ \ \ \ \ \ \ \ \ \ \ \ \ \ \ \ \ }^{k+1} }{~\raisebox{1.15ex}{\gyoung(_4{s}\nabla\nabla_3\cdots\nabla,\  )}} \, ,\ \ \ k=0,1,2,\ldots\,,\ee
will give the shift symmetric $[s,1]$ fields.  We summarize these partially massless and shift symmetric values in figure \ref{s1figure}.  As a final illustrative example, the fields of type $[s,3,2]$ are shown in figure \ref{s32figure}.

\section{Conclusions and Discussion}

We have found the points of enhanced shift symmetry for arbitrary bosonic $p$-forms and mixed symmetry tensors on (A)dS$_D$ space.  This extends \cite{Bonifacio:2018zex}, which covered only the symmetric tensors.  The results for the masses are summarized in \eqref{gensymshsmassee}, and the form of the shift symmetries is \eqref{embeddingbpgeneralpe}.   These fields capture the longitudinal mode that decouples when a massive field approaches a PM point where the top row is activated, so that the gauge symmetry is irreducible.  They also appear, though not as dynamical longitudinal modes, as the endpoints of gauge-for-gauge reducibility chains of reducible PM points.

The shift symmetric fields occur at certain integer values of the dual conformal dimension given by \eqref{genmixwsconfde}. 
These values are all above the unitarity bound $\Delta\geq d+s_1-h_1-1$ for a mixed symmetry state \cite{Metsaev:1997nj,Brink:2000ag,Costa:2014rya}, where $h_1$ is the height of the top ``block" of the tableau, i.e. the number of rows of length $s_1$.  This indicates that the shift fields, in the ordinary quantization, are all unitary on AdS and irreducible with no further null states.   

 On dS, the shift symmetric fields correspond to shortened irreducible representations of the de Sitter algebra (of type ${\cal V}$, in the notation of \cite{Sun:2021thf}) but they generally lie beyond the complementary series and do not correspond to any discrete series points \cite{doi:10.1063/1.1665471,Basile:2016aen}, so they are non-unitary.  A notable exception is the shift symmetric scalars \cite{Folacci:1992xc,Shaynkman:2000ts,Bros:2010wa,Epstein:2014jaa,Chekmenev:2015kzf}, which are unitary in dS$_D$ and correspond to scalar exceptional series representations \cite{vilenkin1978special,Joung:2007je,Gazeau:2010mn,Sun:2021thf}.  
 There are some other lower dimensional exceptions as well, related to the scalars by duality.  For example in $D=3$ the level $k$ shift symmetric $2$ form is dual to the level $k+1$ shift symmetric scalar (as can be seen from the fact that their masses are equal) so the 2-form shift fields are unitary in dS$_3$.  
 
In fact, all the discussions in this paper must be understood modulo these massive dualities.  For example, the construction in Section \ref{2formsection} of the shift symmetric 2-forms as longitudinal modes of a hook field fails in $D=3$ since hook fields are non-dynamical for $D<4$.   But nothing is missed in this case because the shift symmetric 2-form fields are dual to shift symmetric scalars, and these can be constructed as longitudinal modes of massive symmetric tensor fields which are dynamical in $D=3$.  More generally, for low enough dimension where the parent PM field is non-dynamical but the shift symmetric field is, the construction of the shift field as a longitudinal mode fails and the shift field will typically be equivalent by duality to a different shift field whose parent PM field is dynamical.  But this is not always the case, for example in $D=2$ the shift symmetric scalars for $k\geq1$ cannot be constructed as longitudinal modes of any dynamical massive field.

A natural question is whether non-trivial shift symmetric interactions can be found for the more general representations studied here.  
Interactions can always be written using powers of the shift invariant field strength \eqref{genmixsyfise}, so here `non-trivial' means that the interactions are not simply powers of the field strength.  The $k=1,2$ scalars can be given non-trivial self-interactions
 \cite{Goon:2011qf,Goon:2011uw,Bonifacio:2018zex,Bonifacio:2021mrf}, as can the $k=0$ vector \cite{DeRham:2018axr,Bonifacio:2019hrj}, but no other examples are currently known.  
 
 Whether interactions are non-trivial in the sense mentioned above is also tied to whether there are there are non-trivial algebras that could underlie the symmetries \cite{Bogers:2018zeg,Roest:2019oiw}.  A trivial interaction will not deform the abelian algebra of shift symmetries present in the linear theory, whereas a non-trivial interaction should deform the algebra into a non-abelian algebra.  It would be interesting to know if there are finite algebras of the type studied in~\cite{Joung:2015jza} (which are finite subalgebras of higher spin algebras underlying PM Vasiliev theories \cite{Bekaert:2013zya,Basile:2014wua,Alkalaev:2014nsa,Joung:2015jza,Brust:2016zns}), that could be candidates to underly non-trivially interacting shift symmetric $p$-form or mixed symmetry theories.  
 
Non-trivial theories would also presumably have a flat space limit which gives interacting $p$-form or mixed symmetry theories in flat space.  From the point of view of the $S$-matrix, the non-trivial effective field theories on flat space should be theories with enhanced soft limits that allow for a recursive reconstruction of the amplitudes~\cite{Cheung:2014dqa,Cheung:2015ota,Cheung:2016drk,Padilla:2016mno}, so it would be interesting to study if there are such possibilities for fields in these other representations.  In flat space, interactions for $p$-form galileons are known \cite{Deffayet:2010zh,Deffayet:2016von,Deffayet:2017eqq}, though these are presumably Wess-Zumino-like \cite{Goon:2012dy} interactions that do not deform the basic underlying shift symmetries.  It would be interesting to see whether these interactions could be extended to (A)dS and/or deformed into non-trivial interactions, or whether they can harbor hidden special galileon-like non-trivial enhancements.

\vspace{-5pt}
\paragraph{Acknowledgments:}

The author would like to thank James Bonifacio and Austin Joyce for comments on the draft, and acknowledges support from DOE grant DE-SC0009946 and Simons Foundation Award Number 658908.

\bibliographystyle{utphys}
\addcontentsline{toc}{section}{References}
\bibliography{mixshiftdraft-JHEP-2}

\end{document}